\documentclass[twocolumn,showpacs,preprintnumbers,amsmath,amssymb]{revtex4}
\usepackage[dvips]{graphicx}
\usepackage{graphics}
\usepackage{epsf}
\usepackage[dvips]{color}
\input{colordvi.sty}
\begin{document}

\newcommand{\e}[1]{\times 10^{#1}}

\title{Fast kinetic Monte Carlo simulation of strained heteroepitaxy
  in three dimensions}

\author{Chi-Hang Lam and M.T. Lung}
\affiliation{
Department of Applied Physics, Hong Kong Polytechnic University,
Hung Hom, Hong Kong, China}

\author{Leonard M. Sander} 
\affiliation{Michigan Center for Theoretical Physics,
Department of Physics, Randall Laboratory, University of Michigan, Ann
Arbor, MI 48109-1120, USA }

\date{\today}

\begin{abstract}
Accelerated algorithms for simulating the morphological evolution of
strained heteroeptiaxy based on a ball and spring lattice model in
three dimensions are explained. We derive exact Green's function
formalisms for boundary values in the associated lattice elasticity
problems. The computational efficiency is further enhanced by using a
superparticle surface coarsening approximation. Atomic hoppings
simulating surface diffusion are sampled using a multi-step
acceptance-rejection algorithm. It utilizes quick estimates of the
atomic elastic energies from extensively tabulated values modulated by
the local strain. A parameter controls the compromise between accuracy
and efficiency of the acceptance-rejection algorithm.
\end{abstract}

\maketitle

Epitaxial growth techniques enable deposition of a dislocation-free
thin film on a substrate of a different material with a mismatched
lattice constant. For film-substrate combinations such as Ge/Si,
InAs/GaAs and InAs/InP, arrays of three-dimensional (3D) coherent
islands self-assemble spontaneously beyond certain film thicknesses
under appropriate growth conditions \cite{Shchukin, Politi, Freund, Stangl}.
These studies are of much current interest since they are expected to
find applications in future microelectronic devices.
 
One of the most widely studied examples is Ge/Si(100) with a 4\%
lattice misfit. Relatively flat islands called pre-pyramids start to
emerge at 3 monolayers (MLs) of Ge coverage \cite{Mo, Ross,
Vailionis}. Upon further deposition, they quickly grow into truncated
pyramids bounded by four (105) facet planes on the sides and
subsequently into fully grown pyramids which are also called hut
islands. Upon still further deposition, they become dome islands
bounded mainly by (113) facet planes. Finally, large dislocated
islands appear. For the closely related alloy variant
Si$_{1-x}$Ge$_x$/Si(100) with a generally lower $4x$\% misfit, the
development is rather similar and goes through stages characterized by
ripples, hut islands, dome islands and finally dislocated islands
\cite{Floro,Tromp,Sutter}. The structures are however larger and each
transition is postponed to occur at a larger film thickness. The
islands are also more closely packed.

Islands self-assemble because they can relieve the elastic stress in
the heteroepitaxial films. There is theory for island formation that emphasizes nucleation. It suggests that islands
must overcome an energy barrier associated with a critical size so that
the elastic energy gained can balance the extra surface
energy cost \cite{Tersoff_LeGoues}. The theory is reasonably consistent with
experiments at relatively high misfit. At low misfit, the critical
island size and hence the energy barrier are expected to increase and
make nucleation prohibitively difficult. Island
formation mechanisms which do not have a barrier then offer a more promising
explanation. For example, according to the Asaro-Tiller-Grinfeld (ATG) theory
\cite{Asaro,Srolovitz,Spencer93}, strained surfaces are unstable at
sufficiently long wavelengths. Therefore, shallow ripples first
develop from small random initial perturbations and then into
islands. For this mechanism to operate the temperature must be above the surface roughening
transition point. This seems to be the case for the (100) plane of a Si$_{1-x}$Ge$_x$ which is not
a true facet at the experimentally relevant temperatures
\cite{Tersoff02,Rastelli}. 

Direct simulation of the growth of a strained film is  
much more challenging computationally than for the unstrained case
because of the long-range nature of elastic interactions. Early
kinetic Monte Carlo simulations in two dimensions (2D) based on
ball and spring lattice models by Orr et. al \cite{Orr}, Barabasi
\cite{Barabasi}, and Khor and Das Sarma \cite{Khor} have successfully
observed island formation on strained films at sufficiently high
misfits. The mechanism is consistent with the nucleation picture.
These simulations focus on the high misfit regime because islands can
then be small enough to be computationally manageable.  More recently,
much enhanced computational efficiency has been achieved using an
exact Green's function method for the elasticity problem of surface
atoms.  Using also a superparticle coarsening approximation and a
hopping acceptance-rejection sampling method, a much wider range of
morphologies for both high and low temperature regimes characterized
respectively by instability and nucleation roughening mechanisms in
both 2D \cite{Lam,Gray} and 3D \cite{Lung,Lam07} have been explored.
Another advanced approach for the elasticity problem based on the fast
Fourier transform and a multi-grid method applied respectively to the
substrate and film has also been derived \cite{Russo, Russo07}.  All of
these studies work for general film morphologies. Alternatively, one can
limit the study to the evolution of only shallow structures and solve
the elastic problem using the half-space Green's function in the small-slope approximation \cite{Meixner,Zhu}.

Other  computational approaches offer different
compromises between accurate representation of the physics and
computational efficiency. For instance, continuum simulations using
driven diffusion equations with the elastic energy term obtained using
finite element or similar methods are computationally less
intensive. The ATG instability is readily demonstrated \cite{Yang} and
effects of surface anisotropy \cite{Zhang} as well as island
coarsening
\cite{Liu} can be studied in 3D. The importance of alloy
segregation \cite{Tu} and formation of more complicated structures
\cite{ZHuang} have also been studied recently.
Nevertheless, fluctuations in the atomic scale and lattice
discreteness are not accounted for and hence island nucleation and
growth at sub-roughening transition temperatures cannot be studied. Also, off-lattice models are more realistic and island
formation can be simulated at moderate length scales in 2D\cite{Much}.
Moreover, molecular models using realistic semiconductor atomic
potentials have provided better understanding of island stress
distribution \cite{Maxim} and facet structures \cite{Retford} in static
calculations.  Finally, first principles calculations focus on the
statics of fewer atoms but are able to provide important estimates for
the surface energies of relevant facets bounding the islands
\cite{Fujikawa,Shklyaev,Lu}.

Thus kinetic Monte Carlo methods based on lattice models are
unique in allowing large scale dynamic simulations for studying
properties for which atomic discreteness is important while fine
details of the atomic potential and surface energies are of limited
qualitative impacts. In the following, we will explain in detail the
algorithms used in Ref. \cite{Lung} which has enabled some of the
most efficient lattice based kinetic simulations in 3D reported in the
literature.

\section{Model}

The model parameters of our 3D ball and spring model for strained
heteroepitaxy are appropriate to the Si$_{1-x}$Ge$_x$/Si(001) system.
It is based on a cubic lattice with a substrate lattice constant
$a_s=2.715$\AA ~ so that $a_s^3$ gives the correct atomic volume in
crystalline silicon.  The lattice constant $a_f$ of the film material
is related to the lattice misfit $\epsilon=(a_f-a_s)/a_f$ which has a
compositional dependence $\epsilon=0.04 x$.
Nearest and next nearest neighboring atoms are directly connected by
linear elastic springs with force constants $k_1=2 eV/a_s^2$ and
$k_{2}=k_1$ respectively. The shear moduli in (100) and (110)
directions are given by $G_{100}=k_2/a_s$ and
$G_{110}=(k_1+k_2)/2a_s$. They are equal for our choice of $k_1$ and
$k_2$ and this leads to better isotropy of our system.  The elastic
couplings of adatoms with the rest of the system are weak and are
completely neglected for better computational efficiency.
Solid-on-solid conditions and atomic steps limited to at most one atom
high are assumed.  Every topmost atom in the film can hop to a
different random topmost site within a neighborhood of $l\times l$
columns with equal probability. We put $l=17$.
Decreasing the hopping range does not alter our results
significantly. The hopping rate $\Gamma_m$ of a topmost atom
$m$ follows an Arrhenius form:
\begin{equation}
\label{rate}
\Gamma_m = 
{R_0}\exp \left[ -\frac{n_{1m} \gamma_1 + n_{2m} \gamma_2
- \Delta E_m - E_0}{k_{B}T}\right]
\end{equation}
Here, $n_{1m}$ and $n_{2m}$ are the numbers of nearest and next nearest
neighbors of atom $m$. We take $\gamma_1=0.085eV$ and
$\gamma_2=\gamma_1/2$. Single-layer terrace edges along (100) and
(110) directions have energies per unit length given by $(\gamma_1 +
2\gamma_2)/a_s$ and $\sqrt{2} (\gamma_1 + \gamma_2)/a_s$ which are
close to each other.  The energy $\Delta E_m$ is the difference in the
strain energy $E_s$ of the whole lattice at mechanical equilibrium
when the site is occupied versus unoccupied. 
We put $E_0=0.415$eV and $R_0=2D_0/(\sigma a_s)^2$ with $D_0=3.83\times
10^{13}\mbox{\AA}^2 s^{-1}$ and $\sigma^2 =l^2/6$. 
This gives the
appropriate adatom diffusion coefficient for silicon (100)
\cite{Savage}. Note that the hopping rates defined in Eq. (\ref{rate})
following Ref. \cite{Orr} obey detailed balance in contrast to
those adopted in Refs.  \cite{Barabasi,Khor,Russo}. While not
necessarily essential in non-equilibrium situations considered here,
detailed balance adds to the reliability of our results.

The main numerical challenge of our simulation lies on the repeated
calculations of the elastic energies $\Delta E_m$ of topmost atoms in
order to compute the hopping rates $\Gamma_m$ from
Eq. (\ref{rate}). The elasticity problem can be formulated as follows.
First, we note that the strain in a flat film is homogeneous
\cite{Politi}.  This provides a convenient reference position with
displacement $\vec{u}_i=0$ for every atom $i$. From Hooke's law after
applying linearization, the force on atom $i$ exerted by a directly
connected neighbor $j$ is
\begin{equation}
\label{fij}
\vec{f}_{ij} = - {\bf K}_{ij} ( \vec{u}_i -
\vec{u}_j ) + \vec{b}_{ij} 
\end{equation}
where the symmetric matrix ${\bf K}_{ij}=k_{ij} \hat{n}_{ij}
\hat{n}_{ij}^t $ is the modulus tensor and $\vec{b}_{ij} =
(l^0_{ij}-l_{ij}) {\bf K}_{ij} \hat{n}_{ij}$ arises from the
homogeneous stress in flat films. The spring constant $k_{ij}$ equals
either $k_1$ or $k_{2}$ for tangential or diagonal connections
respectively. The unit column vector $\hat{n}_{ij}$ points from the
unstrained lattice position of atom $j$ towards that of atom $i$ and
$t$ denotes transpose.  Furthermore, $l^0_{ij}$ and $l_{ij}$ are
respectively the natural and homogeneously strained spring lengths
which follow easily from $a_s$ and $\epsilon$.  Mechanical equilibrium
requires $\sum_j \vec{f}_{ij}=0$ for each atom $i$. This leads to a
large set of equations coupling the displacements $\vec{u}_i$ of $all$
the atoms. Conventionally, this large set of equations is solved
directly using approximation schemes \cite{Orr, Khor}.  The
solution then gives the elastic energy stored in every spring and
their sum gives the total elastic energy $E_s$ of the whole system in
mechanical equilibrium. Calculating $E_s$ twice with and without the
atom $m$ and comparing the values give $\Delta E_m$. This has to
be done in principle for every topmost atom $m$ after every successful
atomic hop involving non-adatoms. Fast algorithms are hence
indispensable.

Our Green's function method to be explained in the next section is a variant of the
boundary integral method; it directly calculates the displacement of  the
surface atoms only. 
We show here that this is already sufficient for
calculating the elastic energy $E_s$ of the whole system.
First, a surface atom is defined as one which has at least one spring
missing. There are hence more surface atoms than topmost atoms.
Consider virtual forces which adiabatically increase from 0 to
$\sum_{j'} \vec{b}_{jj'}$ applied to every surface atom $j$ where $j'$
is summed over all missing atoms which could have been directly
connected to 
$j$. These forces push all atoms from their mechanical equilibrium
positions to their homogeneously strained lattice positions, i.e. a
displacement $-\vec{u}_j$. The virtual work done, $W>0$, is given by
\begin{equation}
W=- 1/2 \sum_{jj'} \vec{b}_{jj'} \cdot \vec{u}_j
\end{equation}
It equals the difference $E_s^0 - E_s$. Here, $E_s^0$
denotes the strain energy of the homogeneously strained lattice and can be
straightforwardly computed by simple bond counting. Therefore, the
elastic energy $E_s$ of the whole system is given by
\begin{equation}
\label{Es}
E_s = E_s^{0} + \frac{1}{2}\sum_{jj'} \vec{b}_{jj'} \cdot \vec{u}_j  
\end{equation}
which depends only on the positions of the surface atoms \cite{Typo}.

\section{Exact Green's function methods}

We now explain in detail the Green's function method which reduces the
elasticity problem into one involving explicit consideration of only
the surface atoms. This leads to a greatly reduced set of equations
and dramatically reduces the computational burden. It is analogous to
boundary integral methods \cite{Paris}. Either full-space or
half-space Green's functions defined on an extended lattice with
regular boundaries can be used. More importantly, it is exact for
arbitrary morphologies.  This is in sharp contrast to  half-space
Green's functions in the small slope approximation which are often applied
\cite{Tersoff_LeGoues,Meixner,Zhu}.

First, as a mathematical construct, we enlarge the lattice
representing arbitrary morphology by adding $ghost$ atoms with similar
elastic properties to form an extended lattice with regular
boundaries. Unphysical couplings are hence introduced but can be
exactly cancelled. The precise method to achieve this is not unique.
We will first explain a displacement-based Green's function method
which has been applied in our simulations \cite{Lam, Gray, Lung}. For
completeness, we will also introduce closely related force-based
and hypersingular displacement-based Green's function formalisms. They
are not currently adopted but have distinct properties which may lead
to other applications in the future. For all these formalisms, we take
periodic boundary conditions in lateral directions and fixed boundary
conditions at the bottom of the substrate as dictated by the lattice
model. For the top layer in the extended lattice, we adopt fixed
boundary conditions but free boundary conditions are equally good since
all their influences will be exactly cancelled anyway.

\subsection{Displacement-based formalism}
\label{Sec-disp}

A real surface atom $j$ experiences an unphysical force
$\vec{f}_{jj'}$ by a directly connected ghost atom $j$ given by
Eq. (\ref{fij}). To cancel all unphysical forces by the ghost atoms, 
we apply an exactly opposite external force 
\begin{equation}
\vec{f}^e_j=-\sum_{j'}
\vec{f}_{jj'}
\end{equation}
to atom $j$ where the sum is over all ghost atom $j'$ directly
connected to the real atom $j$.  Noting that the elastic properties of the
real and ghost atoms are identical, we apply Eq. (\ref{fij}) and
obtain
\begin{eqnarray}
\label{freal1}
\vec{f}^e_j &=& \sum_{j'}[{\bf K}_{jj'}(\vec{u}_j - \vec{u}_{j'}) -\vec{b}_{jj'}]
\end{eqnarray}
This introduces the unknown displacements $\vec{u}_{j'}$
for the ghost atoms. In this formalism, instead of solving for their
values, we apply another external force
$\vec{f}^e_{j'}$ on every ghost surface atom $j'$ which pushes all ghost
atoms back to the homogeneously strained positions with
$\vec{u}_{j'}\equiv 0$. The force $\vec{f}^e_{j'}$ consists of a term $-\sum_{j}
\vec{f}_{j'j}$ which exactly cancels the spring forces by the real
atoms $j$ as well as a term $\sum_{j} \vec{b}_{j'j}$ which provides the necessary
homogeneous stress. Using Eq. (\ref{fij}) and
$\vec{u}_{j'}\equiv 0$, we get
\begin{eqnarray}
\label{fghost2}
\vec{f}^e_{j'} &=& - \sum_{j} {\bf K}_{jj'} \vec{u}_j 
\end{eqnarray}
where $j$ is summed over all real surface atoms directly connected to
$j'$.
Equation (\ref{freal1}) also reduces to
\begin{eqnarray}
\label{freal2}
\vec{f}^e_j &=& \sum_{j'}({\bf K}_{jj'}\vec{u}_j  -\vec{b}_{jj'})
\end{eqnarray}
With the external forces 
$\vec{f}^e_{j}$ and 
$\vec{f}^e_{j'}$ on the real and ghost surface atoms respectively, the
real lattice is exactly decoupled from the ghost atoms. Therefore, the
problem defined on the extended lattice with the applied forces is
identical to the original physical one.
It is hence legitimate to calculate the displacement $\vec{u}_i$
of any real atom $i$ based on the extended lattice being acted on by the
external forces, i.e.
\begin{equation}
\label{uife}
\vec{u}_i = \sum_j {\bf G}_{ij} \vec{f}^e_j
+ \sum_{j'} {\bf G}_{ij'} \vec{f}_{j'}^e
\end{equation}
where ${\bf G}$ denotes the lattice Green's function for the $extended$
lattice. An important point is that ${\bf G}$ is {\it independent of the film
morphology} and can be computed numerically prior to the start of the
simulation. Substituting Eqs.  (\ref{fghost2}) and (\ref{freal2}) into
Eq. (\ref{uife}), we arrive at the main equation of the
displacement-based Green's function approach
\begin{equation}
\label{ui_new}
\vec{u}_i = 
\sum_{jj'} 
[({\bf G}_{ij} - {\bf G}_{ij'})
{\bf K}_{jj'} \vec{u}_j - {\bf G}_{ij} \vec{b}_{jj'}]
\end{equation}
where the double sum is over all pairs of directly connected real and ghost
surface atoms $j$ and $j'$ respectively. Although Eq. (\ref{ui_new})
holds for any atom $i$, restricting $i$ to only real surface atoms,
Eq. (\ref{ui_new}) now constitutes a greatly reduced set of equations
from which the values of $\vec{u}_i$ for all surface atoms are obtained
numerically using the bi-conjugate gradient method. The elastic energy
$E_s$ of the whole system then follows immediately from
Eq. (\ref{Es}).

\begin{figure}[htp]
\centerline{\epsfxsize 0.7\columnwidth \epsfbox{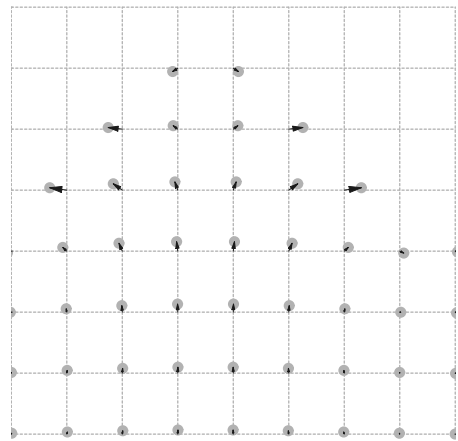}}
\caption{
  \label{Fig-disp}
  Cross-section of a strained film in 3D with a rough
  surface from a small scale simulation. Each arrow represents the
  displacement vector $\vec{u}_i$ of an atom $i$ from its
  homogeneously strained lattice position indicated by the dotted
  lines.  }
\end{figure}

Our simulations do not require calculating the displacements of the
bulk atoms, which can be a time consuming task. Nevertheless, to obtain
$\vec{u}_i$ for a bulk atom $i$ for consistency checks or for the purpose of presentation,
we simply apply Eq. (\ref{ui_new}) and substitute the
displacements of the surface atoms to its R.H.S. Figure \ref{Fig-disp}
shows the calculated displacements of both surface and bulk atoms in
the cross-section of a film from
a small scale 3D simulation.
Another interesting
property of Eq. (\ref{ui_new}) is that the long-range nature of the
elastic interactions is reflected in the coupling coefficient $({\bf
  G}_{ij} - {\bf G}_{ij'}){\bf K}_{jj'}$ which decays as $1/r^2$ in 3D
at
large $r$ where $r$ is the
distance between atoms $i$ and $j$. Therefore, the coupling of the
displacements of the surface atoms is also  long-ranged as expected
and should not be casually truncated. They can be efficiently
accounted for using a coarsening scheme discussed later. This exact
Green's function approach has been motivated by lattice patching and
bond breaking considerations. The derivation is particularly simple
and intuitive. We have since learned that it is a lattice analog of the
boundary integral method for continuum elasticity \cite{Paris}.
It is also closely related to an
earlier result derived algebraically using Dyson's equation \cite{Tewary}.

\subsection{Force-based formalism}
We now explain a closely related but distinct Green's function
approach for the elastic problem which has yet been applied in our
simulations.  The displacements for the ghost surface atoms
$\vec{u}_{j'}$ in Eq.  (\ref{freal1}) are put to zero by applying an
additional set of external forces in the previous discussion.
Alternatively, they can be formally solved without resorting to new
forces. Equation (\ref{freal1}) is first rewritten as
\begin{equation}
\label{fe-force}
\vec{f}^e_i = \sum_{i'}[{\bf K}_{ii'}(\vec{u}_i - \vec{u}_{i'}) -\vec{b}_{ii'}]
\end{equation}
where the sum is over all ghost atoms $i'$ directly connected to the
real atom $i$. Now, $\vec{u}_{i'}$ no longer vanishes in general.
Similar to Eq. (\ref{uife}), the displacements of both real and ghost
atoms $i$ and $i'$ can be expressed using the only external
forces $\vec{f}^e_j$ now present and the Green's
function ${\bf G}$ for the extended lattice as follows
\begin{subequations}
\begin{eqnarray}
\label{fe-ui}
\vec{u}_i &=& \sum_j {\bf G}_{i j} \vec{f}^e_j\\
\vec{u}_{i'} &=& \sum_j {\bf G}_{i' j} \vec{f}^e_j
\end{eqnarray}
\end{subequations}
where the sums are over all real surface
atoms $j$. Substituting the displacements into
Eq. (\ref{fe-force}) gives 
\begin{equation}
\label{force-fe}
\vec{f}^e_i = \sum_{ji'} {\bf K}_{ii'}  
(G_{ij} - G_{i'j}) \vec{f}^e_j
- \sum_{i'} \vec{b}_{ii'}
\end{equation}
This is the main set of equations in the  force-based formalism. They are 
to be solved numerically to obtain the external forces $\vec{f}^e_i$
for all real surface atoms $i$. The displacement $\vec{u}_i$ for the
real surface atoms and hence the elastic energy $E_s$ can then be readily
calculated using Eqs. (\ref{fe-ui}) and (\ref{Es}) respectively.

The coupling coefficient in Eq. (\ref{force-fe}) also decays as
$1/r^2$ with the distance $r$ as in the previous displacement-based
case. For the whole procedure in calculating $E_s$, from both
operation counting and actual numerical implementation, the
computational efficiency of this force-based formalism is similar to
the previous displacement-based formalism. However, the main variable
$\vec{f}^e_i$ here is not a physical quantity and apparently admits no
intuitive coarsening scheme, in contrast to that for the
displacement-based case to be discussed in Section \ref{Sec-SP}. Thus we
have  not pursued  this approach further. We are not aware of
any similar formalism reported in the literature.

\subsection{Hyper-singular displacement-based formalism}

By explicitly demanding mechanical equilibrium of the real surface atoms,
the displacement based formalism in Sec. \ref{Sec-disp} can be further developed 
into another form analogous to the hyper-singular boundary integral
method \cite{Paris}. For a real surface atom $i$, equilibrium implies
\newcommand{\ii}{k}
$\sum_\ii \vec{f}_{i\ii} = 0$
where $\ii$ is summed over all directly connected real neighbors
of $i$, which are not necessarily surface atoms.
Applying Eq. (\ref{fij}), we get
\begin{equation}
\sum_\ii [  {\bf K}_{i\ii} ( \vec{u}_i - \vec{u}_\ii) - \vec{b}_{i\ii} ]= 0
\end{equation}
We now apply Eq. (\ref{ui_new}) twice to express both $\vec{u}_i$ and
$\vec{u}_\ii$ in terms of the displacement of the surface atoms and
obtain after simple algebra
\begin{eqnarray}
\label{hyper-u}
\sum_{j j' \ii } 
{\bf K}_{i\ii}
 \left[ ( {\bf G}_{ij}-{\bf G}_{ij'}) - ( {\bf G}_{\ii j}-{\bf G}_{\ii j'}) \right]
{\bf K}_{jj'} 
\vec{u}_j
\nonumber\\
=
\sum_\ii \vec{b}_{i\ii} 
+\sum_{j j' \ii } 
{\bf K}_{i\ii}
( {\bf G}_{ij}-{\bf G}_{\ii j}) \vec{b}_{jj'} 
\end{eqnarray}
Solution of this set of equations gives $\vec{u}_j$ for all surface
atoms $j$. However, the coefficients and constants in the equations require more
floating point operations to compute and from actual implementation
the equations also take more bi-conjugate gradient steps to
solve. This formalism hence is not adopted. On the other hand, the
coupling constant in Eq. (\ref{hyper-u}) decays as $1/r^3$ in 3D with the
distance $r$ between the corresponding particles. It would be interesting to
explore if the steeper decay can lead to a coarsening
scheme more efficient than the one to be presented next based on the
standard displacement-based formalism.

\section{Superparticle coarsening scheme}
\label{Sec-SP}
The reduced set of equations (Eq. (\ref{ui_new})) from the exact
displacement-based Green's function method for obtaining the
displacements of the surface atoms and hence the elastic energy greatly
speeds up the computation. Yet, it can still be substantially improved
using superparticle coarsening approaches used in Refs.
\cite{Lam,Gray,Lung,Lam07}. The particular implementation adopted in 
Refs. \cite{Lung,Lam07} will be summarized and more details can be
found in Ref. \cite{Lam07}. First, note that the sum over directly
connected pairs of real and ghost surface atoms in
Eqs. (\ref{ui_new}) is
equivalent to the sums over broken bonds of real surface atoms. We
hence rewrite Eq. (\ref{ui_new}) as
\begin{equation}
\label{ui_beta}
\vec{u}_i = 
\sum_{j\beta} \xi_{j\beta}
\left [\Delta {\bf G}_{ij\beta}
{\bf K}_{\beta} \vec{u}_j - {\bf G}_{ij} \vec{b}_{\beta}\right]
\end{equation}
where the sum is now over all real surface atoms $j$ and each
bond direction $\beta$. We put $\xi_{j\beta}=1$ if the
$\beta$th bond of atom $j$ is broken and it is 0 otherwise.  Also,
$\Delta G_{ij\beta}= G_{ij} - G_{ij'}$, $K_\beta=K_{jj'}$ and
$\vec{b}_\beta=\vec{b}_{jj'}$ where the $\beta$th bond of atom $j$
is connected directly to atom $j'$.

The idea of the superparticle approach is as follows. Finding the
strain energy $\Delta E_m$ of atom $m$ needed in Eq. (\ref{rate})
requires calculating the strain energy $E_s$ of the whole lattice
twice with and without atom $m$. Certain fine details of the
surface far away are obviously unimportant and can be neglected.
Specifically, surface atoms are grouped into sets called
superparticles with the $I$th of them denoted by $\Omega_I$. We
neglect fluctuations within a superparticle by assuming identical
displacement $\vec{u}_i\equiv \vec{u}_I$ for each member $i\in
\Omega_I$. Equation (\ref{ui_beta}) can then be approximated by 
\begin{equation}
\label{Coarsen1}
\vec{u}_{I} 
= \sum_J 
\left[ {\bf A}_{IJ} \vec{u}_J - {\vec{B}}_{IJ} \right ] 
\end{equation}
where
\begin{subequations}
\label{Coarsen1AB}
\begin{eqnarray}
\label{Coarsen1A}
{\bf A}_{IJ} &=& 
   \sum_{j\in \Omega_J,\beta} \xi_{j\beta}
   \Delta{\bf G}_{Ij\beta}{\bf K}_{\beta} \\
\label{Coarsen1B}
{\vec{B}}_{IJ} &=& 
   \sum_{j\in \Omega_J,\beta}
   \xi_{j\beta}{\bf G}_{Ij} \vec{b}_{\beta}
\end{eqnarray}
\end{subequations}
The index $I$ in ${\bf G}_{Ij}$ and $\Delta{\bf G}_{Ij\beta}$ denotes
the lattice point closest to the centroid of the superparticle
$\Omega_I$. At this point, individual constituent particles within
the superparticles are still referenced explicitly. To further save
computation time, we 
approximate ${\bf A}_{IJ}$ and ${\vec{B}_{IJ}}$ by 
\begin{subequations}
\label{Coarsen2AB}
\begin{eqnarray}
\label{Coarsen2A}
{\bf A}_{IJ} &=& 
   \sum_{\beta} n_{J\beta} 
   \Delta {\bf G}_{IJ\beta} 
   {\bf K}_{\beta} \\
\label{Coarsen2B}
{\vec{B}}_{IJ} &=& 
   \sum_{\beta} n_{J\beta} {\bf G}_{IJ} \vec{b}_{\beta}
\end{eqnarray}
\end{subequations}
respectively. The Green's functions are now completely indexed by
the centroid positions $I$ and $J$ while
\begin{equation}
n_{J\beta} = 
\sum_{j\in \Omega_J} \xi_{j\beta}
\end{equation}
is the number of broken bonds in superparticle $J$ in direction
$\beta$. However, Eqs. (\ref{Coarsen2AB}) provide good approximations
only when the superparticles $I$ and $J$ are far apart. We still have
to use the more accurate Eqs. (\ref{Coarsen1AB}) for superparticles
close to each other.  In addition, since the problem is always solved
in pairs with identical morphology except for one atom $m$, terms 
in Eqs. (\ref{Coarsen1AB}) and (\ref{Coarsen2AB}) which remain
unchanged in the second calculation are properly re-used. 
In contrast to
conventional methods, the computation time is dominated by the
calculations of the coefficients and constants  in
Eq. (\ref{Coarsen1}) using Eqs.
(\ref{Coarsen1AB}) and (\ref{Coarsen2AB}) rather than the subsequent
iterative solution of the resulting system of equations.  At a given coarseness, the number of superparticles scales
up roughly as $\log L$ for a substrate with a lateral width $L$ for
both 2D and 3D simulations. The computation time which depends on the
number of coefficients and constants in Eq. (\ref{Coarsen1}) thus scales roughly
as $(\log L)^2$.  Further details including the grouping of surface
atoms into superparticles can be found in Ref. \cite{Lam07}.

\section{Atomic hopping acceptance-rejection algorithm}

Even with the Green's function method and the superparticle
approximation, the calculations of the elastic energies of the
potentially hopping atoms are still by far the most time consuming
parts in the kinetic simulations. We adopt a multi-step
hopping algorithm aiming at minimizing the number of these
calculations. It can utilize justifiable approximations to improve
the computational speed. Yet, by tuning a single parameter $\Lambda$
to be defined below, it smoothly crosses over and converges back
to the original exact hopping algorithm. Therefore, the desired
compromise between accuracy and speed can be easily selected to suit
a given set of physical conditions. More importantly, impacts of
any approximations in the hopping algorithm can be easily
accessed by repeating the simulations again with better accuracies.

In the model, each topmost atom is in general allowed to hop to
another column simulating the surface diffusion process. The
conventional approach is to calculate the hopping rate $\Gamma_m$ at
each column $m$ given by Eq.  (\ref{rate}) and then the hopping atom
can be randomly sampled according to the associated probabilities.
After each successful hopping event, the precise surface configuration
changes. Due to the long-range nature of the elastic interactions,
$\Delta E_m$ and $\Gamma_m$ for the whole surface in general also
change and have to be recomputed. In practice, the elastic
interactions are often truncated to a very limited range and thus only
values at a small neighborhood require updating \cite{Orr,Khor}.

We adopt an alternative acceptance-rejection scheme for efficient
sampling of hopping atoms without truncating the elastic interactions.
\newcommand{\ub}{{\Omega_m^+}}
\newcommand{\lb}{{\Omega_m^-}}
First, as will be explained later, easily computable upper and lower
bounds $\ub$ and $\lb$ of $\Delta E_m$ are available. Equation
(\ref{rate}) then gives an upper bound
\begin{equation}
\label{Gamma_bound}
\Gamma_m^{+} = 
{R_0}\exp \left[ 
-\frac{n_{1m} \gamma_1 + n_{2m} \gamma_2 - \ub -
E_0}{k_{B}T}
\right]
\end{equation}
of the rate $\Gamma_m$. All values involved are available and
$\Gamma_m^{+}$ for the whole surface can be easily kept up-to-date.
They are stored in a standard binary tree data structure. At each
Monte Carlo step, we first sample $m$ using $\Gamma_m^{+}$ as the
relative probability efficiently from the binary tree. Since the upper
bound $\Gamma_m^{+}$ instead of the true rate $\Gamma_m$ is used,
atom $m$ hence chosen will hop only with an event acceptance
probability
\begin{equation}
\label{pm}
p_m = \frac{\Gamma_m}{\Gamma_m^{+}} = \exp \left[ 
-\frac{\ub - \Delta E_m}{k_{B}T}
\right]
\end{equation}
It appears that we then need to conduct the intensive computation of
$\Delta E_m$ to find $p_m$, but we can do better. A lower bound
$ p_m^{-}$ of $p_m$ can easily be obtained from
\begin{equation}
\label{p_bound}
 p_m^{-} = \exp \left[ 
-\frac{\ub - \lb}{k_{B}T}
\right]
\end{equation}
Let $\xi$ be an independent uniform random deviate in $[0,1)$. If $\xi
< p_m^{-}$, the hopping event is accepted immediately. An explicit
calculation of $\Delta E_m$ is avoided and this leads to a considerable 
speed up of the simulation. Otherwise, we finally compute
$\Delta E_m$ as described in the previous sections in order to
calculate $p_m$ in Eq. (\ref{pm}). Using the same random number $\xi$, the event is
accepted if $\xi < p_m$. Otherwise, it is rejected.  It can be easily
shown that this acceptance-rejection scheme gives the atomic hopping
rate in Eq. (\ref{rate}) noting that the time elapsed in a Monte
Carlo step is
\begin{equation}
\Delta t = 1/\sum_m  \Gamma_m^{+}
\end{equation}

To calculate the upper and lower bounds required above, we use  
quick estimates $\Omega_m$ of the elastic energies $\Delta
E_m$ which will be explained in Sec. \ref{Sec-QuickEstimates}. With
an estimate $\Omega_m$, we put
\begin{subequations}
\label{bounds}
\begin{eqnarray}
\ub &=& \Omega_m + c^{+}\\
\lb &=& \Omega_m + c^{-} 
\end{eqnarray}
\end{subequations}
where $c^{+}$ and $c^{-}$ are dynamically calculated global
biases. Whenever a calculated value of $\Delta E_m$ does not lie
within the predicted upper bound $\ub$ by a comfortable safety margin
$\Lambda$ taken here as $\Lambda=0.01eV$, $c^{+}$ will be
increased considerably to attain a more conservative bound. Otherwise,
it is decreased slightly for a more aggressive event acceptance rate.
The algorithm for $c^{-}$ is analogous. We also note that $\Delta
E_m \ge 0$ so that $\lb$ is replaced by 0 if it is negative. For adatoms
with elastic couplings neglected, $\ub = \lb = \Omega_m = 0$. In case
an explicit calculation of $\Delta E_m$ is conducted but the hop is
finally rejected, we put $\ub=\lb=\Delta E_m$ until the next
successful hopping event occurs at a neighborhood of $m$. In
particular, for a large safety margin $\Lambda$, the algorithm reduces
into the original exact one in which all values of $\Delta E_m$ has to be
explicitly calculated for all topmost atoms after each successful
hopping event involving a non-adatom. For small $\Lambda$, the
algorithm is more efficient but under or over sampling of certain
hopping events may occasionally occur.

Our acceptance-rejection hopping algorithm is a multi-step one with a
few rather complex components targeting at a good numerical
efficiency. It is further complicated by additional special rules such
as neglecting the elastic interactions of adatoms and forbidding
large atomic steps. Reliable software implementation is
non-trivial because minor coding mistakes affecting only certain
surface configurations often do not lead to disastrous impacts on the
resulting morphology and can be extremely difficult to spot. Hence, we
devote great efforts to guarantee a reliable implementation. One
particularly helpful consistency check is a Boltzmann's distribution
test. Since our model follows detailed balance, an equilibrium surface
follows the Boltzmann's distribution. We perform long test simulations
of annealing in small lattices with all but two atoms frozen. We make
sure that all or most of the not too many possible configurations will
be visited many times. For each configuration, the combined duration
should agree with the Maxwell-Boltzmann's distribution within the
statistical error bar. We repeat the test with frozen atoms arranged
in a wide range of configurations to make sure that hopping events in
all circumstances are simulated with the correct probabilities.

\section{Quick estimates of elastic energies}
\label{Sec-QuickEstimates}
Our hopping algorithms presented in the last section requires an easily
computable estimate $\Omega_m$ for the elastic energy $\Delta E_m$ of
each topmost atom $m$. We now explain our algorithm for generating
$\Omega_m$. Consider first simulations in
2D \cite{Gray}.  Due to
the assumption of linear elasticity, $\Delta E_m
\propto
\epsilon^2$. We write
\begin{equation}
\label{DEm_exact}
\Delta E_m = \epsilon^2 {\Phi_m\left(\{h_i\}_{i=1 .. L}\right) }
\end{equation}
where $\Phi_m$ is the elastic energy of an atom $m$ extrapolated to
unit lattice misfit and it depends on the detailed surface
configuration $\{h_i\}_{i=1 .. L}$ non-trivially.  With long-range
elastic interactions, $\Delta E_m$ and hence $\Phi_m$ depend in
general on the morphology of the whole surface. We split the surface
into a local region $\mathcal L$ centered at $m$ and a distant region
$\mathcal D$ which will be treated accurately and approximately
respectively.
Given a specific local surface configuration defined in $\mathcal L$,
$\Delta E_m$ does not only depend on $\epsilon$ but also significantly
on the morphology in $\mathcal D$. For example, $\Delta E_m$ has a
much larger magnitude when $\mathcal L$ is situated at a highly
stressed valley than at a partially relaxed peak. In fact, what is
most relevant is the resulting local strain induced by $\mathcal D$
averaged over $\mathcal L$.  Let $\lambda_m$ be the horizontal
component of this coarsened strain. Analogous to
Eq. (\ref{DEm_exact}), we hence propose an estimate $\Omega_m$ of
$\Delta E_m$ given by
\begin{equation}
\label{DEm_estimate}
\Omega_m = \lambda_m^2 {\Phi \left(\{h_i\}_{i \in \mathcal L}\right) }
\end{equation}
where the combined effects of the lattice misfit $\epsilon$ and the
morphology in $\mathcal D$ are essentially included in $\lambda_m$.
Impacts due to microscopic details in $\mathcal D$ as well as other
strain components are neglected.  
The elastic energy $\Phi$ at unit strain depends non-trivially on the
local surface configuration. The local configuration is uniquely characterized
by a set of surface steps while the absolute surface height itself is
irrelevant.  Restricting surface step-heights within $\pm 2 a_s$ in
the simulations and taking a local region $\mathcal L$ of 9 columns
wide, there are 8 surface steps and the height of each of them takes
one of five allowed values. For each of the $5^8$
resulting local configurations, $\Phi$ is precomputed and tabulated before
the main simulation commences.  Specifically, we have used in the
precomputation lattices of lateral width $L=32$. The film thickness at
column $m$ is $8a_s$ well cleared of the substrate. In the distant
region $\mathcal D$, the thickness equals $16a_s$ uniformly and this
generates considerable compression.  For every local surface
configuration, the elastic energy $\Delta E_m$ is then explicitly
calculated using our Green's function approach and $\Phi$ is given by
$\Delta E_m/\epsilon^2$ which is independent of the misfit $\epsilon$
used in the pre-computation.

The coarse-grained local strain $\lambda_m$ can be estimated easily
during the simulation noting that it mainly depends on structural
features in $\mathcal D$ at a longer length scale and changes
relatively slowly. Simply by inverting Eq. (\ref{DEm_estimate}) and
averaging, we obtain
\begin{equation}
\label{lambda}
\lambda_m^2 = {  \frac{<\Delta {E}_m>}{<\Phi \left(\{h_i\}_{i\in
\mathcal L}\right)>} }
\end{equation}
where $<\cdot>$ denotes averaging over data associated with the three
most recent explicit calculations of $\Delta E_m$. More precisely,
$\lambda_m$ defined here is consistently smaller than those
proportional to the local strain since $\Phi$ is precomputed with the
local surface located at a valley instead of a flat plane. This
however leads to no degradation in the performance as replacing $\Phi$
by $c\Phi$ for all local configurations for any constant $c>0$ leaves
our algorithm invariant.

\begin{figure}[htp]
\centerline{\epsfxsize 0.9\columnwidth \epsfbox{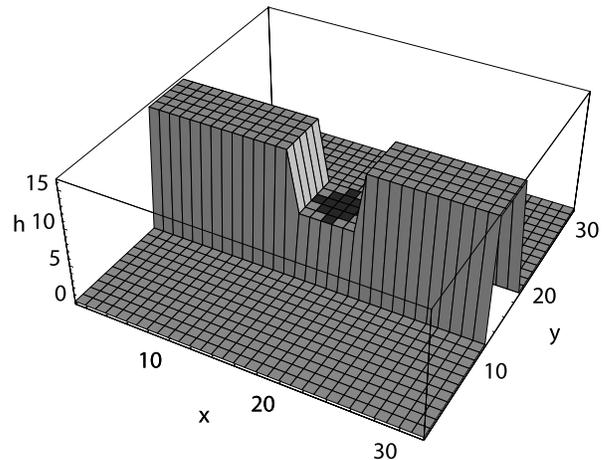}}
\caption{
  \label{Fig-Lconf} Film profile in the distant region $\mathcal D$
for the precomputation of the strain energy table. The local region
$\mathcal L$ is
shaded in black.
}
\end{figure}

For 3D simulations \cite{Lung}, we generalize Eq. (\ref{DEm_estimate}) and approximate
the elastic energy $\Delta E_m$ by
\begin{equation}
\label{DEm_estimate3d}
\Omega_m = 
\lambda_{xm}^2 {\Phi_x( \{ h_i\}_{i\in \mathcal L}) }
+ \lambda_{ym}^2 {\Phi_y(\{ h_i\}_{i\in \mathcal L}) }
\end{equation}
where $\lambda_{xm}$ and $\lambda_{ym}$ are components of the
coarse-grained local strains in planar directions.  The local region
$\mathcal L$ now consists of 13 columns as shown in Fig.
\ref{Fig-Lconf}. Taking into account that step-heights are restricted within $\pm
1$ in our 3D simulations, $3^{12}$ possible surface
configurations characterized by a set of 12 step heights  \cite{Nconf}
are considered. For each configuration, $\Phi_{x}$ is calculated and tabulated using $\Phi_{x}=\Delta
E_m/\epsilon^2$ which follows from Eq. (\ref{DEm_estimate3d})
assuming that the local region is strained only in the $x$ direction
with $\lambda_{xm} \simeq \epsilon$ and $\lambda_{ym} \simeq 0$.
More precisely, the precomputation is based on films on a $32 \times 32 \times 32$
substrate. The frozen morphology in the distant region $\mathcal D$ is
also shown in Fig.
\ref{Fig-Lconf}. It is the simplest morphology which generates a significant
strain only in the $x$ direction. Then $\Phi_{y}$ is obtained from $\Phi_x$ by
symmetry.

In contrast to the 2D case, $\lambda_{xm}$ and $\lambda_{ym}$ cannot
be solved directly but are instead obtained from a simple least square
fit. We define an error measure $\mathcal E_m = ( \Delta E_m - \Omega_m
)^2$. Whenever and explicit calculation of $\Delta E_m$ is conducted,
the local strain components are updated according to the steepest descent
approach as follows
\begin{eqnarray}
\lambda_{xm} \longleftarrow \lambda_{xm} - \eta \frac{\partial \mathcal
E_m}{\partial \lambda_{xm}}\\
\lambda_{ym} \longleftarrow \lambda_{ym} - \eta \frac{\partial \mathcal
E_m}{\partial \lambda_{ym}}
\end{eqnarray}
where the rate constant $\eta$ is taken as 0.5.  Furthermore,
$\lambda_{xm}$ and $\lambda_{ym}$ are relatively smooth functions of
position. To suppress statistical errors, after each steepest descent
step, they are further replaced by weighted averages with values at
neighboring columns, i.e.
\begin{eqnarray}
\lambda_{xm} &\longleftarrow& w \lambda_{xm} + (1-w) <\lambda_{xi}>_m\\
\lambda_{ym} &\longleftarrow& w \lambda_{ym} + (1-w) < \lambda_{yi}
>_m
\end{eqnarray}
where $w=0.8$ and $<...>_m$ denotes averaging over the 8 neighboring
columns of site $m$.

\section{results}

\begin{figure}[htp]
\centerline{\epsfxsize 0.8\columnwidth \epsfbox{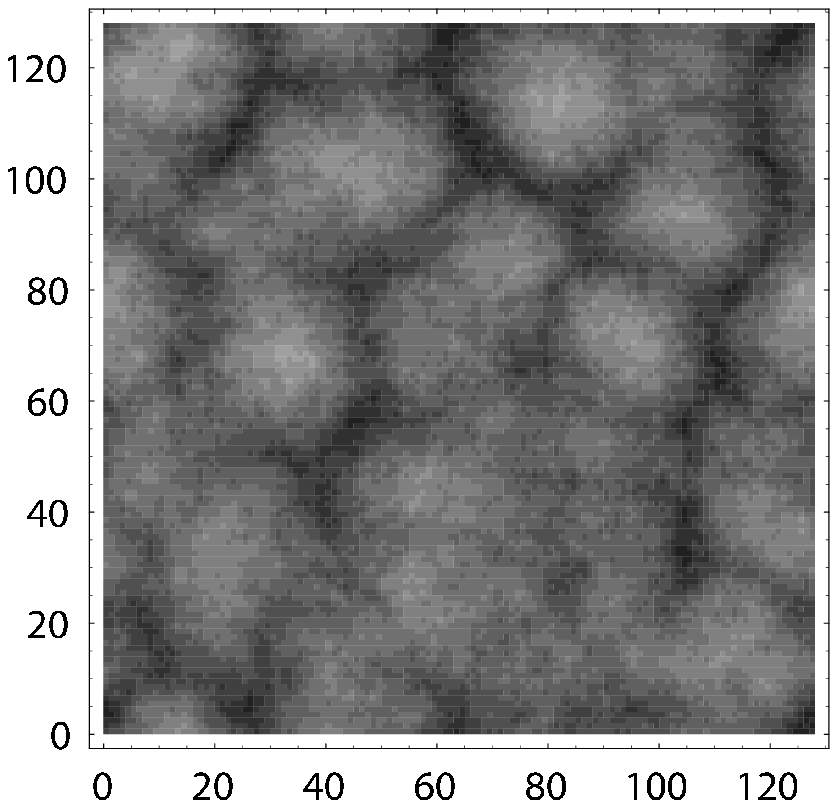}}
\centerline{\epsfxsize 0.9\columnwidth \epsfbox{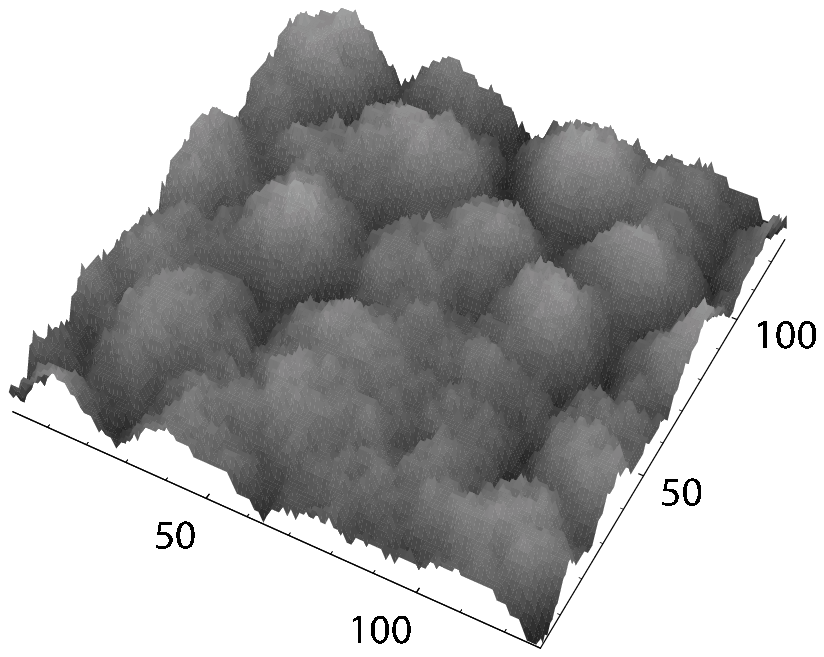}}
\caption{
  \label{Fig-surface} Surface from a simulation of annealing at misfit
  $\epsilon=6\%$ and temperature $T=1000K$ in top view and 3D
view. The peak to peak roughness is 9 monolayers. }
\end{figure}

\begin{figure}[htp]
\centerline{\epsfxsize 0.9\columnwidth \epsfbox{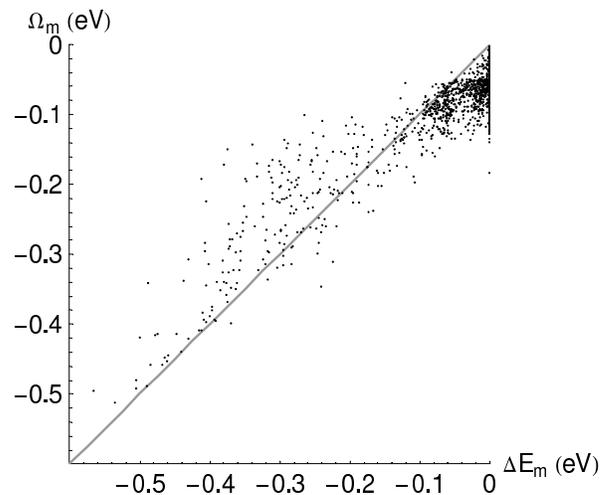}}
\caption{
  \label{Fig-Omega} Plot of $\Omega_m$ against $\Delta E_m$ where $\Delta
  E_m$ is the elastic energy of a surface atom calculated using the
  superparticle approach and $\Omega_m$ is its quick estimate.}  
\end{figure}

Figure \ref{Fig-surface} shows an initially flat film from a typical
simulation after annealing for $66\mu s$ at a lattice misfit
$\epsilon=6\%$ and temperature $T=1000K$. Islands have started to
form and the roughening mechanism is expected to follow the
Asaro-Tiller-Grinfeld instability as explained in
Ref. \cite{Lung}. The nominal thickness is 5 monolayers. The substrate
used has $128 \times 128 \times 64$ lattice sites and this is the
largest considered in similar kinetic simulations reported in the
literature. The pre-computed Green's functions have been calculated
for an extended lattice with $128 \times 128 \times (64+30)$ sites so
that films with a maximum local thickness up to $30$ layers can be
simulated. Totally, we have carried out $6.5\e{6}$ hopping attempts (Monte
Carlo steps) each of which involves a sampling event according to
Eq. (\ref{Gamma_bound}) based on the upper bound $\ub$ of the elastic energy.
About $2.2\e{6}$ of them are successful hops by adatoms with elastic interactions
neglected. Of the remaining $4.3\e{6}$ non-adatom hopping attempts,
$1.8\e{6}$, i.e. 42\% are
accepted directly using Eq. (\ref{p_bound}) based on the lower bound
$\lb$. In the other $2.5\e{6}$ attempts , $\Delta E_m$ is explicitly computed
and a further $0.6\e{6}$ attempts, i.e. 13\% are accepted using Eq. (\ref{pm}).  
Therefore, a combined 55\% of all non-adatom hopping attempts are
accepted while the remaining 45\% are rejected.
Totally, there are hence $4.6\e{6}$ successful hopping events with
$2.4\e{6}$ of them involving non-adatoms.
About 1.04 explicit calculations of $\Delta
E_m$ are carried out for each successful non-adatom hopping event compared
with nearly $128\times 128$ calculations in principle.
Figure
\ref{Fig-Omega} plots the quick estimates $\Omega_m$ against the more
accurate values $\Delta E_m$ obtained using the superparticle
approximation. The pairs of values are sampled randomly throughout the
simulation.
For the whole simulation, $\Delta E_m$ actually lies outside the
predicted bounds only with a small chance of less than $0.8$\%. The
resulting over- or under-samplings of hopping events have shown to be 
negligible in smaller scale simulations. 

The simulations take 18 days to execute on a 2.2GHz Core Duo Pentium
computer. The repeated calculations of $\Delta E_m$ consume about 85\%
of the CPU time. We use 110 superparticles. Each calculation takes
about 0.5s.  About 95\% of that goes to computing the coefficients and
constants in setting up Eq. (\ref{ui_new}). This part of our codes
have been carefully written to implement both parallel processing with
the dual cores and vector processing with arrays each consisting of
four single precision floating point numbers provided by the streaming
SIMD Extensions 2 (SSE2) of the processor. The Intel C++ compiler is
used.  The remaining 5\% of the computation load is spent on the
iterative solution of the equations using the bi-conjugate gradient
method. It uses Intel's mathematical kernel library which also takes
advantage of the parallel and vector facilities. We use a relatively
small tolerance of $0.0001$\% as the convergence condition for the
iterative solution of the superparticle displacements. The overall
error in the calculation of the elastic energy $\Delta E_m$ is
numerically checked to be within about 5\% which is essentially
due to the superparticle approximation only. The run time of a
simulation is expected to scale roughly as $L^2(\log L)^2$ for a
substrate with $L\times L$ surface sites. The factors $L^2$
and $(\log L)^2$ respectively account for the number of hopping surface
particles and the computation load for each elastic energy calculation
as explained in Sec. III. For instance, when we repeat our simulation
using a smaller substrate with $L=64$, it takes about 3 days
to execute.

\section{Discussion}

Algorithms for fast Monte Carlo simulation of the morphological
evolution of strain heteroepitaxial thin films are presented. A
Green's function approach and a superparticle surface coarsening
scheme enable efficient calculations of the elastic energies of film
atoms. Atomic hopping events following rates in the Arrhenius form are
selected using an acceptance-rejection algorithm. The algorithm
utilizes estimates of the elastic energies of topmost atoms easily
computable from tabulated values for similar local surface
configurations after taking into account local strains.  With
these algorithms, kinetic Monte Carlo simulations have been conducted
much more efficiently than it was possible previously.

This work was supported by HK RGC, Grant No. PolyU-5009/06P. LMS is supported by
NSF grant DMS 0553487.


\begin{references}
\bibitem{Nconf} 
Some of the $3^{12}$ local configurations uniquely indexed
by a selected set of 12 steps actually contain forbidden double-steps
and they are assigned null entries in our elastic energy table. This is because there are
indeed 16 surface steps between the 13 columns due to the existence of loops on
a 2D plane. Explicit knowledge of the exact number of valid configurations
however is not needed in our algorithm.
\end{references}
\end{document}